\documentclass[aps,prl,reprint,superscriptaddress,floatfix,nofootinbib,nobibnotes,
               amsmath,amssymb]{revtex4-2}
\usepackage{graphicx}
\usepackage{multirow}
\usepackage{pifont}
\usepackage{xcolor}
\usepackage{soul}

\newcommand{\cmark}{\ding{51}}
\newcommand{\xmark}{\ding{55}}
\makeatletter
\newcommand{\mcpcell}[2]{{%
  \setlength{\fboxsep}{0pt}%
  \hspace*{-\tabcolsep}%
  \colorbox{#1}{\makebox[\dimexpr2em+2\tabcolsep\relax][c]{%
    \@arstrut#2}}%
  \hspace*{-\tabcolsep}%
}}
\makeatother
\newcommand{\mcpcheck}{\mcpcell{green!18}{\cmark}}
\newcommand{\mcpdash}{\mcpcell{red!15}{\mbox{--}}}
\newcommand{\rr}{\raggedright}
\newcommand{\vtier}[4]{\multirow{#1}{*}[#4]{{\fboxsep=1pt%
  \colorbox{black!8}{\makebox[8.5mm]{\rotatebox[origin=c]{90}{%
  \parbox{#2}{\centering\scriptsize\pretolerance=-1\exhyphenpenalty=50\hyphenpenalty=50\textit{#3}}}}}}}}
\usepackage{hyperref}
\hypersetup{colorlinks=true,linkcolor=blue,citecolor=blue,urlcolor=blue}

\newcommand{\reportnumber}{FERMILAB-PUB-26-0631-T}
\usepackage[angle=0,scale=1,color=black,firstpage=true,opacity=1]{background}
\SetBgContents{\footnotesize\sf{\reportnumber}}
\SetBgPosition{current page.north east}
\SetBgHshift{-2in}
\SetBgVshift{-0.5in}

\newcommand{\pkgname}{HEPTAPOD}
\newcommand{\adname}{Diagrammatica}

\begin{document}

\preprint{[preprint number]}

\title{Reining in an Agentic Harness for High Energy Physics}

\author{Tony Menzo}
\email{amenzo@ua.edu}
\affiliation{Department of Physics and Astronomy, University of Alabama, Tuscaloosa, AL 35487, USA}
\affiliation{Fermi National Accelerator Laboratory, Batavia, IL 60510, USA}
\author{George T.~Fleming}
\affiliation{Fermi National Accelerator Laboratory, Batavia, IL 60510, USA}
\author{Konstantin T.~Matchev}
\affiliation{Department of Physics and Astronomy, University of Alabama, Tuscaloosa, AL 35487, USA}
\author{Stephen Mrenna}
\affiliation{Fermi National Accelerator Laboratory, Batavia, IL 60510, USA}
\author{Alexander Roman}
\affiliation{Department of Physics and Astronomy, University of Alabama, Tuscaloosa, AL 35487, USA}

\date{\today}

\begin{abstract}
Agentic systems now address tasks across theoretical, phenomenological, and experimental high energy physics (HEP), but their scientific capabilities remain difficult to reuse across different large language models, providers, and harnesses. We argue that stable parts of these workflows should be promoted into versioned scientific operations and exposed through common protocols. Existing general-purpose harnesses can then be specialized for HEP through task-specific sets of tools and skills, while community-maintained registries would make these capabilities discoverable and citable. We identify mismatches in conventions, assumptions, and domains of validity among independently developed operations as a potential obstacle to their composition, and discuss machine-readable scientific contracts as one possible solution. These design principles and evaluation guidelines provide a near-term path toward a portable and community-maintained agentic harness for HEP.
\end{abstract}

\maketitle


Large language models (LLMs) situated in executable environments act as \emph{agents}; planning, writing and running code, invoking external programs, and checking results across sequential turns.
Given a task, the behavior of an agent depends on both the raw capability of the underlying LLM and the \emph{harness} supplied around it.
The harness is the structured assembly of context-management, state, execution policies (retry, stopping, recovery), logging policies, and external capability access that transforms a general model into a domain-specific agent.  
Existing harnesses for software development specialize general models by supplying context about the operations and conventions involved in writing, testing, and maintaining software.
A domain-specific harness for high energy physics (HEP) must similarly manage HEP context and capabilities under the validation requirements of the field.
Determining how separately specified domain capabilities are presented and composed, which operations are left to the model, and where human verification remains required is a distinct discipline, referred to as \emph{harness engineering}.

Agentic systems now span theoretical, phenomenological, and experimental HEP. 
In Table \ref{tab:harness_landscape}, these systems are classified by the structure of the interface exposed to the agent, \textit{i.e.}~the interaction surface and types of operations the agent has available to complete a given task.
\emph{Improvisational} or \emph{free-form} (token-by-token) systems provide the agent with rich domain context, \emph{e.g.}~retrieved literature~\cite{Moreno:2026mqk,Bakshi:2025fgx,Badea:2026klb}, role-specific instructions~\cite{Plehn:2026gxv,Diefenbacher:2025zzn,Diefenbacher:2026azr}, prescribed workflows~\cite{Gendreau-Distler:2025fsj,Esmail:2026jpb}, or curated knowledge bases~\cite{Miao:2025sms,Tan:2026ier}, while leaving substantial portions of the task implementation to free-form code and command generation.
\emph{Structured-handoff} systems leave scientific operations to agent-generated code and commands, but constrain the intermediate artifacts passed between agents or workflow stages, using, for example, fixed JSON schemas, machine-readable project specifications, or structured analysis plans~\cite{He:2026jjb,Lucente:2026kgh,Birk:2026zpd,Costa:2026oew,Hammad:2026ged,Saito:2026tfq,Gao:2026cpd}.
\emph{Fixed-backend} systems reduce free-form implementation by having the agent generate command-line invocations, configuration files, templates, or DSL programs consumed by fixed software~\cite{ColliderAgent,Jiao:2026dsu,Xiao:2026gal,Guo:2026ifi}.
\emph{Tool-constrained} systems further restrict the operational interface by providing the agent with a library of callable \emph{tools}%
\footnote{The term `tool' is used loosely in the HEP literature, encompassing handbooks and command templates~\cite{ColliderAgent}, CLI utilities~\cite{Faroughy:2026dkj}, domain-specific languages~\cite{Jiao:2026dsu}, packaged agent skills~\cite{Wang:2026cof,Gao:2026cpd,Xiao:2026gal}, and schema-based wrappers around scientific software~\cite{Agrawal:2026lvg,Desai:2026nmx,Hill:2026naa,Saad:2026pan}. Here, following the LLM tool-use literature, we reserve \emph{tool} to refer to a function interface through which a model invokes a computer program external to the model by generating a call and its arguments~\cite{wang2024toolsanywaysurveylanguage,schick2023toolformerlanguagemodelsteach,qin2023toolllmfacilitatinglargelanguage}. Versioning, validation, and schema-constrained generation are additional properties of a tool rather than part of its definition.}%
, which are exposed through bounded function interfaces often accompanied by specialized context (typically referred to as  \emph{skills})~\cite{Menzo:2025cim,Menzo:2026qrl,Agrawal:2026lvg,Desai:2026nmx,Hill:2026naa,Saad:2026pan,Wang:2026jjn,Gao:2026cpd,Lugato:2026osi}.
In both fixed-backend and tool constrained systems, scientific operations are validated before being made available to the agent. 
In contrast to free-form and structured-handoff systems, whose agent-generated implementations must be validated as they are produced, this prior validation reduces the remaining burden of establishing scientific correctness to verifying correct invocation, configuration, and composition. 
Tool-constrained systems differ further in when interface conformance is enforced, with \emph{valid-by-rejection} systems checking calls after generation and \emph{valid-by-construction} systems constraining generation itself (Sec.~\ref{sec:versioned_operations}).

\begin{figure*}[t]
\centering
\includegraphics[width=0.7\textwidth]{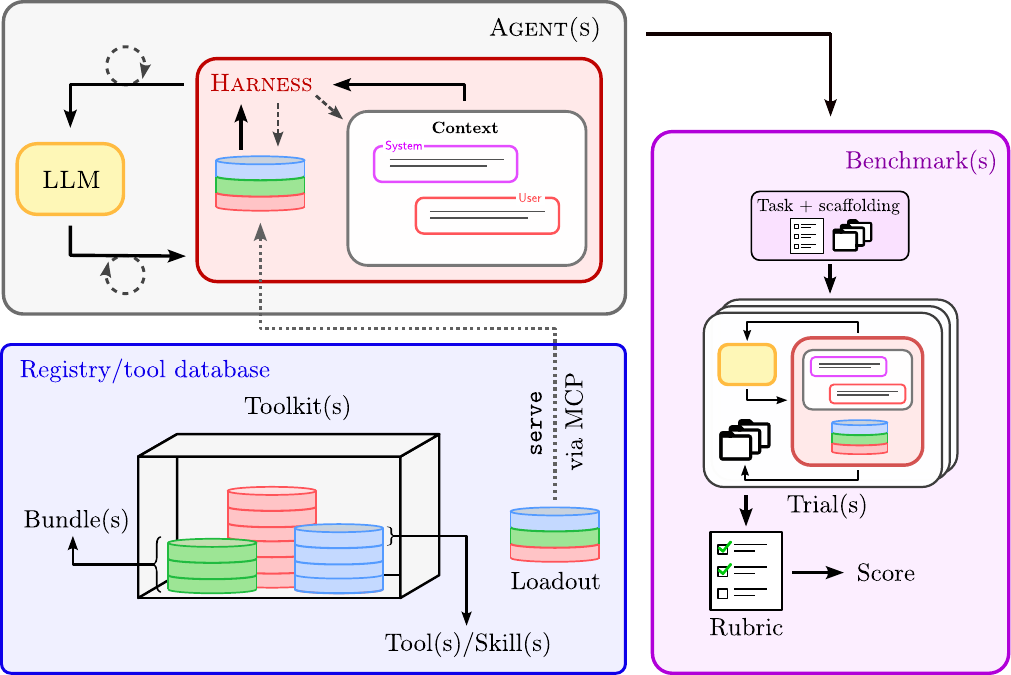}
\caption{Schematic depiction of harness and its accompanying infrastructure.}
\label{fig:stack}
\end{figure*}

To date, most HEP agentic systems are developed as self-contained demonstrations that bind scientific tasks, supplied context, tools, models, providers, and runtimes together (see Fig.~\ref{fig:stack}).
This organization is natural for rapid development and early exploration, but it prevents validated capabilities from becoming reusable scientific infrastructure. 
A capability validated in one system cannot readily be reused in another, and a measured change in performance cannot be cleanly attributed to the model, harness, or tools. 
This uncoordinated approach is particularly consequential for research-level HEP, where each generated implementation must be independently verified, often without a cheap, independent oracle to evaluate correctness. 
Similarly, while developing benchmarks (see Table~\ref{tab:benchmarks}) address increasingly realistic tasks, they generally fix the full system configuration and do not preserve sufficient evidence to reproduce or reinterpret their outcomes. 
Altogether, the field has a growing collection of compelling demonstrations without a common technical and experimental framework through which they can be compared, reused, and built upon.

In this Letter, we outline a set of design principles and supporting infrastructure for a HEP-focused agentic harness, focusing primarily on exploratory research assistance, where validated, modular automation supports more reliable and auditable execution of long-horizon tasks within researcher-directed workflows.
The same design principles also provide a foundation for increasingly autonomous scientific systems as direct human review recedes.
We argue that stable scientific operations should be promoted into validated, versioned, and citable artifacts, while the scientific task, task scaffolding (\textit{i.e.}~the sandbox and workspace initialization, prompts, \textit{etc.}), model and inference controls, inference provider, harness, and capability loadout(s) remain separately resolvable, so that a change in outcome can be attributed to a single factor.
Portability requires the same operation to be discoverable and invokable through different harnesses, resolvable to a citable implementation and environment, and composable with other operations under explicit scientific conventions.
This separation also allows an operation to be held fixed as other components are varied and their effects on the distribution of outcomes are compared under reproducible conditions.
In Sec.~\ref{sec:versioned_operations}, we motivate and define versioned scientific operations as validated, citable artifacts that can be exposed through common protocols and assembled into task-specific loadouts.
In Sec.~\ref{sec:infrastructure}, we describe the registries, scientific compatibility mechanisms, and benchmarking framework needed to make these operations discoverable, composable, and reproducible.
Terminology drawn from software engineering and machine learning is collected in Appendix~\ref{app:glossary}.


\begin{table*}[t]
\centering
\scriptsize
\renewcommand{\arraystretch}{1.15}
\hyphenpenalty=10000 \exhyphenpenalty=10000
\begin{tabular}{ c p{3.15cm} p{5.30cm} c c p{4.35cm} }
\hline\hline
\noalign{\vskip 2.5pt}
& Name & Architecture & MCP & \shortstack{Multi-\\agent} & Domain / role \tabularnewline
\hline
\vtier{16}{48mm}{Free-form}{0pt} & \rr MadAgents~\cite{Plehn:2026gxv}   & \rr Free-form code generation   &   \mcpdash   & \cmark   & \rr Collider phenomenology \tabularnewline
& \rr HEP Data Analysis Agents~\cite{Gendreau-Distler:2025fsj}   & \rr Free-form generation in fixed workflow DAG   &   \mcpdash   & \cmark   & \rr Experimental analysis \tabularnewline
& \rr JFC~\cite{Moreno:2026mqk}   & \rr Free-form generation, literature-grounded   &   \mcpdash   & \cmark   & \rr Experimental analysis \tabularnewline
& \rr CoLLM~\cite{Esmail:2026jpb}   & \rr Free-form generation, structured prompt   &   \mcpdash   & \xmark   & \rr Collider DL analysis \tabularnewline
& \rr PhysMaster~\cite{Miao:2025sms,Tan:2026ier}   & \rr Free-form generation, validated-trace memory   &   \mcpdash   & \cmark   & \rr Theory / lattice QCD \tabularnewline
& \rr Agents of Discovery~\cite{Diefenbacher:2025zzn}   & \rr Free-form code generation   &   \mcpdash   & \cmark   & \rr New-physics analysis \tabularnewline
& \rr ArgoLOOM~\cite{Bakshi:2025fgx}   & \rr Orchestrated module wrappers, RAG-grounded   &   \mcpdash   & \cmark   & \rr Cosmology, collider, nuclear \tabularnewline
& \rr ALEPH thrust~\cite{Badea:2026klb}   & \rr Free-form generation, physicist-directed   &   \mcpdash   & \xmark   & \rr Experimental analysis \tabularnewline
& \rr SFitter Agents~\cite{Diefenbacher:2026azr}   & \rr Orchestrator with consultant agents   &   \mcpdash   & \cmark   & \rr Global SMEFT re-casting \tabularnewline
& \rr CLVisc Agent~\cite{Wang:2026cof}   & \rr Free-form execution, agent-authored skill   &   \mcpdash   & \xmark   & \rr Heavy-ion hydrodynamics \tabularnewline
& \rr Amplitude program search~\cite{Gu:2026zbf}   & \rr Coding agent in generate--evaluate--select loop   &   \mcpdash   & \xmark   & \rr Scattering amplitudes \tabularnewline
& \rr Detector co-design~\cite{Chung:2026elz}   & \rr Agentic loop over bi-level optimizer   &   \mcpdash   & \xmark   & \rr Detector design \tabularnewline
\hline
\vtier{9}{26mm}{Structured handoffs}{0pt} & \rr Dr.\,Sai~\cite{He:2026jjb}   & \rr Role agents over experiment software   &   \mcpdash   & \cmark   & \rr Experimental analysis \tabularnewline
& \rr DarkAgents~\cite{Lucente:2026kgh}   & \rr Sub-agents with JSON handoff contracts   &   \mcpdash   & \cmark   & \rr Astroparticle / dark sectors \tabularnewline
& \rr SHARP~\cite{Birk:2026zpd}   & \rr Machine-readable project spec, subagents   &   \mcpdash   & \cmark   & \rr Analysis reproduction \tabularnewline
& \rr AgentRivet~\cite{Costa:2026oew}   & \rr Provider-agnostic agents, output schemas   &   \mcpdash   & \cmark   & \rr Analysis preservation \tabularnewline
& \rr Semantic differencing~\cite{Hammad:2026ged}   & \rr Multi-agent trace and critique of generated code   &   \mcpdash   & \cmark   & \rr Analysis audit \tabularnewline
& \rr Selection-list codegen~\cite{Saito:2026tfq}   & \rr Structured selection list between stages   &   \mcpdash   & \cmark   & \rr Analysis reproduction \tabularnewline
& \rr LQCDMaster~\cite{Gao:2026cpd}   & \rr Hybrid: skill-guided generation, deterministic Wick tool   &   \mcpdash   & \cmark   & \rr Lattice QCD \tabularnewline
\hline
\vtier{4}{11mm}{Fixed backend}{0pt} & \rr ColliderAgent~\cite{ColliderAgent}   & \rr CLI/template-mediated   &   \mcpcheck   & \cmark   & \rr Collider phenomenology \tabularnewline
& \rr HepScript~\cite{Jiao:2026dsu}   & \rr Constrained-grammar DSL   &   \mcpdash   & \xmark   & \rr Experimental analysis \tabularnewline
& \rr EasyScan\_HEP~2~\cite{Xiao:2026gal}   & \rr Agent-written configuration, fixed backend   &   \mcpdash   & \xmark   & \rr Parameter scans \tabularnewline
& \rr SMEFT-Pheno-Agent~\cite{Guo:2026ifi}   & \rr Locked configuration, phase manifests   &   \mcpdash   & \xmark   & \rr SMEFT phenomenology \tabularnewline
\hline
\vtier{9}{26mm}{Valid-by-rejection}{0pt} & \rr RooAgent~\cite{Desai:2026nmx}   & \rr Typed tool registry over LangGraph and MCP   &   \mcpcheck   & \xmark   & \rr Experimental analysis \tabularnewline
& \rr FERMIACC~\cite{Agrawal:2026lvg}   & \rr Typed schemas over a guarded pipeline   &   \mcpdash   & \cmark   & \rr Particle theory \tabularnewline
& \rr GRACE~\cite{Hill:2026naa}   & \rr Tool-invocation DAG; value-level verification against physics constraints   &   \mcpdash   & \xmark   & \rr Detector / experiment design \tabularnewline
& \rr bsm\_agent~\cite{Saad:2026pan}   & \rr Tool calling over a symbolic backend   &   \mcpdash   & \xmark   & \rr BSM model building \tabularnewline
& \rr LeWRON~\cite{Wang:2026jjn}   & \rr Audited toolbox with auditor agents   &   \mcpdash   & \cmark   & \rr Particle cosmology \tabularnewline
& \rr Archi~\cite{Lugato:2026osi}   & \rr ReAct agent over corpus and live MCP tools   &   \mcpcheck   & \xmark   & \rr Experimental operations \tabularnewline
\hline
\vtier{4}{12mm}{Valid-by-con\-struc\-tion}{0pt} & \rr \pkgname\ (event generation)~\cite{Menzo:2025cim}   & \rr Typed schema constrains decoding   &   \mcpcheck   & \xmark   & \rr Collider phenomenology \tabularnewline
& \rr \pkgname\ (\adname)~\cite{Menzo:2026qrl}   & \rr Typed schema constrains decoding   &   \mcpcheck   & \xmark   & \rr Symbolic computation \tabularnewline
\noalign{\vskip 6pt}
\hline\hline
\end{tabular}
\caption{%
Agentic systems in HEP at the time of writing, grouped by their primary interface. A checkmark on a green background in the MCP column denotes that the system's domain toolset, or a domain backend developed with it, is exported through an MCP server; a dash on a red background denotes that such export is not reported.}
\label{tab:harness_landscape}
\end{table*}


\begin{table*}[t]
\centering
\scriptsize
\renewcommand{\arraystretch}{1.15}
\hyphenpenalty=10000 \exhyphenpenalty=10000
\begin{tabular}{ p{2.35cm} p{4.35cm} p{4.75cm} p{4.35cm} }
\hline\hline
\rr Name  & \rr Scope  & \rr Grading  & \rr Integrity control / retained evidence  \tabularnewline
\hline
\rr Collider-Bench~\cite{Faroughy:2026dkj}  & \rr LHC analysis reproduction; 10 tasks from 4 papers  & \rr Histogram fidelity, no hand-written rubric; 3 repeats; cost reported  & \rr LLM judge audits workspace and session trace for fabrication  \tabularnewline
\rr TPBench~\cite{Chung:2025nsd}  & \rr HEP theory and cosmology; 57 problems  & \rr Automated numeric verification; holistic grading of reasoning  & \rr Public/private split maintained against leakage  \tabularnewline
\rr CritPt~\cite{Zhu:2025qnm}  & \rr Frontier physics research incl.\ HEP; 71 challenges, 190 checkpoints  & \rr Physics-format-aware auto-grading; reliability metric; cost reported  & \rr Unpublished problems, guess-resistant by construction  \tabularnewline
\rr PRL-Bench~\cite{Miao:2026gmx}  & \rr End-to-end physics research; 100 papers across 5 subfields  & \rr Objective verifiability; expert-validated tasks  & \rr Published-source tasks; retained execution evidence not specified  \tabularnewline
\hline\hline
\end{tabular}
\caption{%
Benchmarks relevant to agentic HEP at the time of writing.}
\label{tab:benchmarks}
\end{table*}


\section{Designing Portable Scientific Operations for Agents}\label{sec:versioned_operations}

Computational workflows in HEP routinely compose reusable operations whose implementations, conventions, and domains of validity can be inspected and cited.
These operations are typically distributed and versioned as part of larger codebases containing many modular classes and functions.
In the context of a harness, however, agents may discover, select, and compose individual callable operations drawn from several independently maintained software packages.
To make the resulting workflow reproducible, the implementation of each operation, its agent-facing interface, and any external software it invokes should resolve to fixed versions.
We refer to an operation resolved in this way as a \emph{versioned scientific operation}.

A versioned scientific operation can be exposed to an agent through several interfaces.
The agent may write code against a library API, generate a command-line invocation, configuration, or DSL program for a fixed backend, or invoke a tool that interfaces with a registered function.
When using the native software interfaces, the model expresses the invocation as code, commands, or configuration that the harness then executes.
A tool interface instead presents a bounded operation and its arguments directly to the model, allowing the harness to dispatch the resulting call to the underlying software.
This avoids requiring the agent to inspect large portions of a codebase or reconstruct the syntax and glue code needed for each invocation.
Where the native interface already validates inputs, returns structured outputs, and reports actionable errors, the tool can remain a thin adapter.
Its additional value comes from providing a compact, task-oriented description, narrowing the available interface to scientifically meaningful operations, returning consistent diagnostics, and allowing the harness to record calls, provenance, and artifacts.
The scientific implementation and its validation remain in the underlying software.
Tools can therefore be treated as first-class interfaces for agents, with a toolkit defining the callable operations that the model can select and compose. Within a toolkit, \textit{bundles} group related tools into coherent capabilities, while task-specific loadouts select toolkits, bundles, or individual tools.

Schema-based tool interfaces can enforce conformance at different stages.
Under \emph{valid-by-rejection}~\cite{Desai:2026nmx,Agrawal:2026lvg,Saad:2026pan,Wang:2026jjn}, a call is checked after generation and invalid structure or values are rejected with diagnostics.
Under \emph{valid-by-construction}~\cite{Menzo:2025cim,Menzo:2026qrl,willard2023efficientguidedgenerationlarge,zhang2024dontfinetunedecodesyntax}, constrained decoding enforces the schema during generation when supported by the inference provider.
Complementary value-level checks can reject calls that are well-formed but physically meaningless~\cite{Hill:2026naa}.
While these schema mechanisms reduce invalid tool invocations, they do not establish scientific validity/physically meaningful usage, which still remains grounded in prior tool-validation as well as human-in-the-loop verification of tool composition.

When an operation is exposed as a tool, however, its agent-facing representation also depends on the conventions of the inference provider.
OpenAI and Anthropic, for example, use different representations for tool definitions, call identifiers, arguments, results, and their placement within the conversation~\cite{OpenAIFunctionCalling,AnthropicToolUse}.
Implementing an operation directly against these provider-specific formats would require separate integrations across providers and harnesses.
Community-adopted protocols, such as the Model Context Protocol (MCP), address this portability problem by standardizing tool discovery, machine-readable interface descriptions, invocation, and result exchange~\cite{MCP}.
An MCP-capable harness translates tool descriptions, calls, and results between MCP and the format of the active inference provider, while the corresponding server executes each operation directly or maps it onto an existing library API, command-line program, or DSL.
At the time of writing, MCP is widely supported by major coding harnesses, including Claude Code~\cite{ClaudeCode}, Codex~\cite{OpenAICodex}, and Google Antigravity~\cite{GoogleAntigravity}, as well as provider-agnostic harnesses and agent frameworks such as OpenCode~\cite{OpenCode}, Orchestral AI~\cite{Roman:2026rcn}, and aisuite~\cite{AISuite}.
Among the HEP systems surveyed in Table~\ref{tab:harness_landscape}, however, documented export of domain operations through MCP remains uncommon.
Additionally, given the growing prevalence of multi-agent systems in HEP (see Table~\ref{tab:harness_landscape}), compatibility with protocols such as A2A for communication between independently deployed agents may provide an additional layer of interoperability as this ecosystem continues to develop~\cite{A2AProtocol}.
Related standards address communication with agent clients, user interfaces, and existing APIs~\cite{ACP,AGUI,OpenAPISpec}.
Although the architecture proposed here is not tied to any single protocol, the breadth of current MCP support makes it the practical protocol through which HEP developers can expose scientific operations once and make them available across harnesses and inference providers.
A HEP-focused harness can then be assembled by equipping an existing general-purpose harness with a task-specific loadout of portable domain operations rather than binding those capabilities to a self-contained system.

Tools expose callable operations, while skills supply procedural knowledge about how those operations should be selected, composed, and checked.
Together, a task-specific set of tools and skills forms the capability loadout presented by the harness.
The harness manages context and persistent state, records calls and artifacts, and determines how operations are presented and recovered when they fail.
Provider-agnostic agent interfaces, separately distributed skills, and MCP-exposed scientific operations already provide partial precedents for this architecture~\cite{Costa:2026oew,Xiao:2026gal,Desai:2026nmx}.

The appropriate boundary between free-form generation and versioned scientific operations depends on the maturity of the underlying method.
Free-form generation remains appropriate while a method is evolving or no stable abstraction exists.
As part of a workflow becomes established, it can be implemented, validated, and exposed through a fixed interface, while the agent remains responsible for selecting and composing it with the rest of the calculation.
Domain tools can make established calculations accessible to models that cannot reliably implement them independently~\cite{Menzo:2025cim}.
As models become more capable, the value of these operations extends beyond enabling implementation to making higher-level selections and compositions auditable through a resolved record of calls, versions, and artifacts~\cite{Li:2026agentharness,Li:2026harnesscoevolution}.
Versioned scientific operations therefore standardize the stable components of a workflow without prescribing its scientific trajectory.

\section{Infrastructure for Portable and Reproducible Agentic HEP} \label{sec:infrastructure}

Portable scientific operations become community infrastructure once researchers can install and discover them, assemble scientifically compatible loadouts, and resolve each invocation to the implementation and environment that produced it.
To evaluate the resulting harnesses, a measurement framework is needed that separately identifies the factors influencing agent performance and preserves all relevant evidence from each trial.
Together, these requirements call for installable packages that resolve dependencies, versioned registries and loadouts that support discovery and reuse, mechanisms that declare and check scientific compatibility, and benchmark conditions and trial records that enable reproducible evaluation.

A \emph{versioned registry} indexes domain capabilities by name and version and stores the metadata (\textit{e.g.}~external software configuration, external dependency versions, \textit{etc}.)
required to resolve each registered operation to a fixed implementation and agent-facing interface, with explicit dependencies and a citable version~\cite{SoftwareCitationPrinciples,CWL,REANA}. The official MCP Registry provides a general discovery layer for public MCP servers~\cite{MCPRegistry}. As tools and skills proliferate, presenting all capabilities to every agent becomes impractical because descriptions consume context, overlapping operations complicate selection, and incompatible conventions or dependencies may not compose safely. Researchers and benchmarks can instead define named loadouts that resolve compatible tool and skill versions and record the bounded set exposed to the agent. \textsc{ToolBase} provides one early command-line interface for searching a larger collection of capabilities and curating them into named loadouts~\cite{toolbase}. Registered loadouts serve as executable documentation for published workflows, while version histories preserve earlier configurations and keep updates discoverable.

Reliable composition of registered scientific operations depends on compatible conventions, approximations, and domains of validity. Depending on the workflow, compatibility may involve units, reference frames, event-record semantics, particle and status conventions, weight normalization, perturbative order, PDF and tune choices, detector conditions, object definitions, or statistical conventions. HEP has long addressed this problem through standards and interchange formats that define the semantics of data exchanged between independently developed codes~\cite{Allanach:2008qq,Alwall:2006yp,Buckley:2019xhk}. Existing HEP metadata efforts demonstrate both the importance of recording such information and the difficulty of maintaining it across heterogeneous analysis
systems~\cite{Belov:2010xm,Khoo:2022pja}. 
An agent composing operations developed by different groups may not inspect their implementations or recognize assumptions left implicit in their documentation. This creates a potential obstacle for portable HEP harnesses and motivates a common mechanism for declaring and checking scientific compatibility. Typed workflow descriptions and semantic tool annotations provide useful precedents~\cite{CWL,Kasalica:2021ape}. At the time of writing, no shared scientific type framework provides this compatibility layer for agentic HEP. One possible realization would associate each registered operation with a machine-readable scientific contract. Unlike an interface schema, a scientific contract
describes the scientific properties required of its inputs and guaranteed for its outputs using namespaced and versioned declarations. A compatibility resolver checks global constraints when assembling a loadout and compares required and provided properties at each handoff, classifying the composition as compatible, convertible through a registered adapter, incompatible, or unresolved. Properties fixed by invocation arguments are attached to the resulting artifact and checked at subsequent handoffs. The full contracts could remain outside the model context, while mismatches, available conversions, and unresolved assumptions could be presented to the agent and researcher.

As stated above, existing HEP benchmarks (see Table \ref{tab:benchmarks}) provide increasingly realistic shared tasks and grading procedures, but their reported configurations do not yet resolve every component needed to vary models, providers, harnesses, and capability loadouts independently~\cite{Faroughy:2026dkj,Chung:2025nsd,Zhu:2025qnm,Miao:2026gmx}. In a \emph{science-forward benchmarking framework}, a benchmark condition consists of the scientific task, task scaffolding, model and inference controls, inference provider, harness, and capability loadout, each specified and resolved separately. Together, these parameters influence the distribution of trajectories observed across repeated trials. For each condition, the task fixes the objective, inputs, protected ground truth, required evidence, and grading contract; the scaffolding fixes the initialized sandbox and workspace, prompts, supplied information, and procedural recipes; and the loadout fixes the scientific operations available to the agent. Resolving these factors separately allows a validated operation to be held fixed while models, providers, harness policies, or other components are varied through matched or factorial comparisons. A framework implementing these requirements is in preparation~\cite{Menzo:inprep}.

Repeated trials under a fixed condition sample the distribution of trajectories and produce a corresponding distribution of graded outcomes. The mean outcome measures average performance, while its variance and lower tail characterize reliability; the number of trials and sampling policy must therefore be fixed and reported for each condition. Physics-aware rubrics grade scientific content independently of toolkit or source-code structure, allowing results to be compared across different implementations. For long workflows, stagewise checks of assumptions, intermediate derivations or calculations, simulation and data-processing stages, and final observables can localize failure across theoretical, phenomenological, and experimental tasks. Symbolic answers, numerical tolerances, distributions, and contours require different comparisons with retained evidence.

A complete trial record allows scientific failures to be distinguished from infrastructure failures and supports \emph{post-hoc grading}, in which retained trajectories and artifacts are evaluated or re-evaluated without repeating inference~\cite{ROCrate,W3CPROV,CERNAnalysisPreservation}. Naturally, each resulting score should additionally identify the versioned rubric under which it was assigned. The record includes the requested and resolved configuration, prompts, interface descriptions, trajectory, researcher interventions, artifacts, software versions, grading evidence, termination status, and resource use. Benchmark designers should keep ground truth outside the agent's environment, control capabilities that could expose it, and inspect trajectories for attempted access. Reproducibility means preserving the declared condition, realized environment, and evidence behind the score.

\section{Conclusion}

Agentic systems in HEP have demonstrated useful scientific capabilities, but those capabilities remain difficult to reuse across models, providers, and harnesses. Their reported performance results remain similarly difficult to reproduce or reinterpret from the available evidence. We argue that the barriers to reuse and reproducibility need to be lowered for advances in future systems to contribute to cumulative progress across the field. Drawing on established software-engineering practices, we identify versioned scientific operations and the adoption of common protocols as near-term remedies for this fragmentation. A versioned scientific operation resolves its agent-facing interface, implementation, dependencies, validation record, and external software to a citable artifact. Together with associated skills, these operations can be assembled into task-specific loadouts that specialize existing general-purpose harnesses for HEP without binding scientific capabilities to a single system.

Realizing a portable and community-maintained agentic harness for HEP requires both shared design principles and community infrastructure. Common tool protocols, registries, and client-side loadout curation can build on existing software-engineering systems. We urge developers building agentic systems for HEP to expose the scientific operations used by their systems through versioned, MCP-compatible interfaces and provide the metadata required to resolve each operation to its interface implementation, underlying software, dependencies, and validation record.

Agentic HEP still lacks a shared scientific type framework for declaring and checking the scientific compatibility of composed operations. Machine-readable scientific contracts would provide a concrete route to this layer by declaring the conventions, assumptions, and domains of validity governing each operation. Developing and maintaining the shared vocabulary and compatibility rules for these contracts will require broad support from the HEP community. Existing HEP benchmarks provide shared tasks, but a common standard for resolved benchmark conditions and complete trial records remains to be established. Evaluations should separately resolve the scientific task, task scaffolding, model and inference controls, inference provider, harness, and capability loadout, repeat trials under fixed conditions, and preserve complete trial records. The immediate goal is a portable and community-maintained agentic harness for HEP in which each new system contributes validated scientific operations that future systems can discover, compose, and build upon.

\section*{Acknowledgments}
TM thanks Daniel C.~Hackett for valuable discussions that sharpened the motivation and framing of this work. TM is supported in part by the Shelby Endowment for Distinguished Faculty at the University of Alabama and by Fermilab via Subcontract 725339. 
The work of AR and KM is supported in part by the Shelby Endowment for Distinguished Faculty at the University of Alabama and by Fermilab via Subcontract 731293, in support of DOE Award No.\ DE-SCL0000090 ``HEP AmSC IDA Pilot: Knowledge Extraction'' and DOE Award No.\ DE-SCL0000152 ``USQCD AmSC Infrastructure Provision''. 
This manuscript has been authored by Fermi Forward Discovery Group, LLC under Contract No.~89243024CSC000002 with the U.S. Department of Energy, Office of Science, Office of High Energy Physics. 
This work was performed in part at the Aspen Center for Physics, which is supported by National Science Foundation grant PHY-2210452.

\bibliography{harnessing_agentic_hep}

@misc{Plehn:2026gxv,
    author = "Plehn, Tilman and Schiller, Daniel and Schmal, Nikita",
    title = "{MadAgents}",
    eprint = "2601.21015",
    archivePrefix = "arXiv",
    primaryClass = "hep-ph",
    month = "1",
    year = "2026"
}

@misc{Esmail:2026jpb,
    author = "Esmail, W. and Hammad, A. and Nojiri, M.",
    title = "{CoLLM: AI engineering toolbox for end-to-end deep learning in collider analyses}",
    eprint = "2602.06496",
    archivePrefix = "arXiv",
    primaryClass = "hep-ph",
    month = "2",
    year = "2026"
}

@misc{ColliderAgent,
    author = "Qiu, Shi and Cai, Zeyu and Wei, Jiashen and Li, Zeyu and Yin, Yixuan and Cao, Qing-Hong and Liu, Chang and Luo, Ming-xing and Yuan, Xing-Bo and Zhu, Hua Xing",
    title = "{An End-to-end Architecture for Collider Physics and Beyond}",
    eprint = "2603.14553",
    archivePrefix = "arXiv",
    primaryClass = "hep-ph",
    reportNumber = "CPTNP-2026-012",
    month = "3",
    year = "2026"
}

@inproceedings{Gendreau-Distler:2025fsj,
    author = "Gendreau-Distler, Eli and Ho, Joshua and Kim, Dongwon and Le Pottier, Luc Tomas and Wang, Haichen and Yang, Chengxi",
    title = "{Automating High Energy Physics Data Analysis with LLM-Powered Agents}",
    booktitle = "{39th Annual Conference on Neural Information Processing Systems}: {Includes Machine Learning and the Physical Sciences (ML4PS)}",
    eprint = "2512.07785",
    archivePrefix = "arXiv",
    primaryClass = "physics.data-an",
    month = "12",
    year = "2025"
}

@misc{Moreno:2026mqk,
    author = "Moreno, Eric A. and Bright-Thonney, Samuel and Novak, Andrzej and Garcia, Dolores and Harris, Philip",
    title = "{AI Agents Can Already Autonomously Perform Experimental High Energy Physics}",
    eprint = "2603.20179",
    archivePrefix = "arXiv",
    primaryClass = "hep-ex",
    month = "3",
    year = "2026"
}

@misc{Chung:2026elz,
    author = "Chung, Wonyong and Liu, Qibin and Wu, Liangyu and Gonski, Julia",
    title = "{Agentic-AI Detector Co-design and Optimization in Vertically-Integrated Differentiable Full Simulations}",
    eprint = "2604.21804",
    archivePrefix = "arXiv",
    primaryClass = "physics.ins-det",
    month = "4",
    year = "2026"
}

@misc{Agrawal:2026lvg,
    author = "Agrawal, Prateek and Craig, Nathaniel and Madden, Amalia and Lombera, I{\~n}igo Valenzuela",
    title = "{The FERMIACC: Agents for Particle Theory}",
    eprint = "2603.22538",
    archivePrefix = "arXiv",
    primaryClass = "hep-ph",
    month = "3",
    year = "2026"
}

@misc{Miao:2025sms,
    author = "Miao, Tingjia and others",
    title = "{PhysMaster: Building an Autonomous AI Physicist for Theoretical and Computational Physics Research}",
    eprint = "2512.19799",
    archivePrefix = "arXiv",
    primaryClass = "cs.AI",
    month = "12",
    year = "2025"
}

@misc{Bakshi:2025fgx,
    author = "Bakshi, S. D. and others",
    title = "{ArgoLOOM: agentic AI for fundamental physics from quarks to cosmos}",
    eprint = "2510.02426",
    archivePrefix = "arXiv",
    primaryClass = "hep-ph",
    reportNumber = "ANL-199516",
    month = "10",
    year = "2025"
}

@misc{Menzo:2025cim,
    author = {Menzo, Tony and Roman, Alexander and Gleyzer, Sergei and Matchev, Konstantin and Fleming, George T. and H{\"o}che, Stefan and Mrenna, Stephen and Shyamsundar, Prasanth},
    title = "{HEPTAPOD: Orchestrating High Energy Physics Workflows Towards Autonomous Agency}",
    eprint = "2512.15867",
    archivePrefix = "arXiv",
    primaryClass = "hep-ph",
    reportNumber = "FERMILAB-PUB-25-0923-CSAID-ETD-T",
    month = "12",
    year = "2025"
}

@misc{Menzo:2026qrl,
    author = "Menzo, Tony and Roman, Alexander and Fleming, George T. and Gleyzer, Sergei and Matchev, Konstantin T. and Mrenna, Stephen",
    title = "{Agentic Diagrammatica: Towards Autonomous Symbolic Computation in High Energy Physics}",
    eprint = "2603.26990",
    archivePrefix = "arXiv",
    primaryClass = "hep-ph",
    reportNumber = "FERMILAB-PUB-26-0208-T",
    month = "3",
    year = "2026"
}

@misc{Li:2026agentharness,
    author       = "Li, Junjie and Xiao, Xi and Zhang, Yunbei and Liu, Chen and Zhao, Lin and Liao, Xiaoying and Ji, Yingrui and Wang, Janet and Gu, Jianyang and Ge, Yingqiang and Xu, Weijie and Fang, Xi and Xu, Xiang and Zhao, Tianchen and Kim, Youngeun and Wang, Tianyang and Hamm, Jihun and Krishnaswamy, Smita and Huan, Jun and Reddy, Chandan",
    title        = "{Agent Harness Engineering: A Survey}",
    howpublished = "\url{https://openreview.net/pdf?id=eONq7FdiHa}",
    year         = "2026",
    note         = "OpenReview preprint, accessed 24 August 2026"
}

@misc{Li:2026harnesscoevolution,
    author       = "Li, Jinzhe and Wu, Yuan and Chang, Yi",
    title        = "{Harness Engineering for LLM Agents: A Survey of Harness Component Taxonomy, Evaluation, and Model--Harness Coevolution}",
    howpublished = "\url{https://www.preprints.org/manuscript/202606.2203}",
    doi          = "10.20944/preprints202606.2203.v1",
    year         = "2026",
    note         = "Preprints.org preprint, accessed 24 August 2026"
}

@misc{Diefenbacher:2025zzn,
    author = {Diefenbacher, Sascha and Hallin, Anna and Kasieczka, Gregor and Kr{\"a}mer, Michael and Lauscher, Anne and Lukas, Tim},
    title = "{Agents of Discovery}",
    eprint = "2509.08535",
    archivePrefix = "arXiv",
    primaryClass = "hep-ph",
    month = "9",
    year = "2025"
}

@misc{He:2026jjb,
    author = "He, Mingfeng and others",
    title = "{Dr.Sai: An agentic AI for real-world physics analysis at BESIII}",
    eprint = "2604.22541",
    archivePrefix = "arXiv",
    primaryClass = "hep-ex",
    month = "4",
    year = "2026"
}

@misc{Desai:2026nmx,
    author = "Desai, Aman",
    title = "{RooAgent: An LLM Agent for Root-Based High Energy Physics Analysis}",
    eprint = "2605.17318",
    archivePrefix = "arXiv",
    primaryClass = "hep-ph",
    month = "5",
    year = "2026"
}

@misc{Diefenbacher:2026azr,
    author = "Diefenbacher, Sascha and Plehn, Tilman and Schiller, Daniel and Schmal, Nikita",
    title = "{Agentic Re-Casting using Agentic Re-Simulations}",
    eprint = "2607.22813",
    archivePrefix = "arXiv",
    primaryClass = "hep-ph",
    month = "7",
    year = "2026"
}

@misc{Lucente:2026kgh,
    author = "Lucente, Michele and Pascoli, Silvia and Sala, Filippo and Zandi, Matteo",
    title = "{DarkAgents}",
    eprint = "2606.11157",
    archivePrefix = "arXiv",
    primaryClass = "hep-ph",
    month = "6",
    year = "2026"
}

@misc{Badea:2026klb,
    author = "Badea, Anthony and Chen, Yi and Maggi, Marcello and Lee, Yen-Jie",
    collaboration = "Electron-Positron Alliance",
    title = "{Agentic AI -- Physicist Collaboration in Experimental Particle Physics: A Proof-of-Concept Measurement with LEP Open Data}",
    eprint = "2603.05735",
    archivePrefix = "arXiv",
    primaryClass = "hep-ex",
    month = "3",
    year = "2026"
}

@misc{Birk:2026zpd,
    author = "Birk, Joschka and Kasieczka, Gregor and Mishra-Sharma, Siddharth and Nachman, Benjamin and Noll, Dennis and Wamorkar, Tanvi",
    title = "{A Scientific Human-Agent Reproduction Pipeline}",
    eprint = "2604.18752",
    archivePrefix = "arXiv",
    primaryClass = "hep-ph",
    doi = "10.5281/zenodo.21078068",
    month = "4",
    year = "2026"
}

@misc{Costa:2026oew,
    author = {Costa, Antonio J. and Doglioni, Caterina and G{\"u}tschow, Christian and Pilkington, Andrew D. and Sinha, Sukanya},
    title = "{AgentRivet: an automated system for producing Rivet routines from journal publications}",
    eprint = "2606.13535",
    archivePrefix = "arXiv",
    primaryClass = "hep-ex",
    reportNumber = "MCNET-26-14",
    month = "6",
    year = "2026"
}

@misc{Wang:2026cof,
    author = "Wang, Qi and Pang, Long-Gang and Pu, Shi and Wang, Xin-Nian",
    title = "{CLVisc Agent for autonomous relativistic hydrodynamics studies}",
    eprint = "2607.27822",
    archivePrefix = "arXiv",
    primaryClass = "nucl-th",
    month = "7",
    year = "2026"
}

@misc{Guo:2026ifi,
    author = "Guo, Yu-Chen and Wang, Jie and Yang, Ji-Chong",
    title = "{SMEFT-Pheno-Agent: a natural-language-driven AI agent for machine-learning-assisted Standard Model Effective Field Theory phenomenology}",
    eprint = "2607.22331",
    archivePrefix = "arXiv",
    primaryClass = "hep-ph",
    month = "7",
    year = "2026"
}

@misc{Lugato:2026osi,
    author = "Lugato, Pietro and others",
    title = "{Archi: Agentic Operations at the CMS Experiment}",
    eprint = "2606.04755",
    archivePrefix = "arXiv",
    primaryClass = "hep-ex",
    reportNumber = "FERMILAB-PUB-26-0400-CMS-PPD",
    month = "6",
    year = "2026"
}

@misc{Gao:2026cpd,
    author = "Gao, Haofei and others",
    title = "{LQCDMaster: Agentic Scientific Computing for Lattice Quantum Chromodynamics Research}",
    eprint = "2607.15001",
    archivePrefix = "arXiv",
    primaryClass = "hep-lat",
    month = "7",
    year = "2026"
}

@misc{Tan:2026ier,
    author = "Tan, Jin-Xin and Miao, Ting-Jia and Zhang, Mu-Hua and Pang, Xiang-He and Liu, Ze-Xi and Zhang, Lin-Feng and Chen, Si-Heng and Wang, Wei",
    title = "{Automated Extraction of Collins-Soper Kernel from Lattice QCD using An Autonomous AI Physicist System}",
    eprint = "2603.22471",
    archivePrefix = "arXiv",
    primaryClass = "hep-lat",
    month = "3",
    year = "2026"
}

@misc{Gu:2026zbf,
    author = "Gu, Yi and Krippendorf, Sven",
    title = "{Scattering Amplitudes as Programs: Self-Evolving Search for Theory and Event Generation}",
    eprint = "2607.21629",
    archivePrefix = "arXiv",
    primaryClass = "hep-ph",
    month = "7",
    year = "2026"
}

@misc{Hammad:2026ged,
    author = "Hammad, Ahmed and Nojiri, Mihoko",
    title = "{Articulating Assumptions in AI-Generated Scientific Analyses through Task Decomposition}",
    eprint = "2607.05762",
    archivePrefix = "arXiv",
    primaryClass = "cs.SE",
    month = "7",
    year = "2026"
}

@inproceedings{Saito:2026tfq,
    author = "Saito, Masahiko and Kishimoto, Tomoe and Tanaka, Junichi",
    title = "{Development of an LLM-Based System for Automatic Code Generation from HEP Publications}",
    eprint = "2604.14696",
    archivePrefix = "arXiv",
    primaryClass = "physics.data-an",
    month = "4",
    year = "2026"
}

@misc{Wang:2026jjn,
    author = "Wang, Isaac R.",
    title = "{LeWRON: Agentic Analysis of Electroweak Phase Transitions}",
    eprint = "2606.19425",
    archivePrefix = "arXiv",
    primaryClass = "hep-ph",
    reportNumber = "FERMILAB-PUB-26-0408-T",
    month = "6",
    year = "2026"
}

@misc{Saad:2026pan,
    author = "Saad, Shaikh",
    title = "{Large Language Model-Assisted Framework for BSM Model Building}",
    eprint = "2606.21316",
    archivePrefix = "arXiv",
    primaryClass = "hep-ph",
    month = "6",
    year = "2026"
}

@misc{Jiao:2026dsu,
    author = "Jiao, Junkun and Liu, Tong and Li, Ke and Song, Weimin and Liao, Yipu and Zhang, Bolun and Liu, Beijiang and Yuan, Chang-Zheng and Sun, Yue",
    title = "{HepScript: A Dual-Use DSL for Human-AI Collaborative Data Analysis Workflows in High-Energy Physics}",
    eprint = "2605.01423",
    archivePrefix = "arXiv",
    primaryClass = "hep-ex",
    month = "5",
    year = "2026"
}

@misc{Xiao:2026gal,
    author = "Xiao, Yang and Yue, Yuanfang and Zhang, Yang",
    title = "{EasyScan{\_}HEP 2: Agent-Ready Parameter Scans for High-Energy Physics}",
    eprint = "2606.31214",
    archivePrefix = "arXiv",
    primaryClass = "hep-ph",
    month = "6",
    year = "2026"
}

@inproceedings{Hill:2026naa,
    author = "Hill, Justin and Ryoo, Hong Joo",
    title = "{GRACE: an Agentic AI for Particle Physics Experiment Design and Simulation}",
    eprint = "2602.15039",
    archivePrefix = "arXiv",
    primaryClass = "hep-ex",
    month = "1",
    year = "2026"
}

@misc{Faroughy:2026dkj,
    author = "Faroughy, Darius A. and Palacios Schweitzer, Sofia and Pang, Ian and Mishra-Sharma, Siddharth and Shih, David",
    title = "{Collider-Bench: Benchmarking AI Agents with Particle Physics Analysis Reproduction}",
    eprint = "2605.13950",
    archivePrefix = "arXiv",
    primaryClass = "cs.LG",
    month = "5",
    year = "2026"
}

@article{Chung:2025nsd,
    author = {Chung, Daniel J. H. and Gao, Zhiqi and Kvasiuk, Yurii and Li, Tianyi and M{\"u}nchmeyer, Moritz and Rudolph, Maja and Sala, Frederic and Tadepalli, Sai Chaitanya},
    title = "{Theoretical physics benchmark (TPBench){\textemdash}a dataset and study of AI reasoning capabilities in theoretical physics}",
    eprint = "2502.15815",
    archivePrefix = "arXiv",
    primaryClass = "cs.LG",
    doi = "10.1088/2632-2153/adfcb0",
    journal = "Mach. Learn. Sci. Tech.",
    volume = "6",
    number = "3",
    pages = "030505",
    year = "2025"
}

@misc{Zhu:2025qnm,
    author = "Zhu, Minhui and others",
    title = "{Probing the Critical Point (CritPt) of AI Reasoning: a Frontier Physics Research Benchmark}",
    eprint = "2509.26574",
    archivePrefix = "arXiv",
    primaryClass = "cs.AI",
    month = "9",
    year = "2025"
}

@misc{Miao:2026gmx,
    author = "Miao, Tingjia and others",
    title = "{PRL-Bench: A Comprehensive Benchmark Evaluating LLMs' Capabilities in Frontier Physics Research}",
    eprint = "2604.15411",
    archivePrefix = "arXiv",
    primaryClass = "cs.LG",
    month = "4",
    year = "2026"
}

@misc{zhang2024dontfinetunedecodesyntax,
      title={Don't Fine-Tune, Decode: Syntax Error-Free Tool Use via Constrained Decoding}, 
      author={Kexun Zhang and Hongqiao Chen and Lei Li and William Wang},
      year={2024},
      eprint={2310.07075},
      archivePrefix={arXiv},
      primaryClass={cs.CL},
      url={https://arxiv.org/abs/2310.07075}, 
}

@misc{MCP,
    author       = "{Anthropic}",
    title        = "{Model Context Protocol}",
    howpublished = "\url{https://modelcontextprotocol.io}",
    year         = "2024",
    note         = "Accessed 2026"
}

@misc{OpenAIFunctionCalling,
    author       = "{OpenAI}",
    title        = "{Function Calling}",
    howpublished = "\url{https://developers.openai.com/api/docs/guides/function-calling}",
    year         = "2026",
    note         = "Accessed 24 August 2026"
}

@misc{AnthropicToolUse,
    author       = "{Anthropic}",
    title        = "{Tool Use with Claude}",
    howpublished = "\url{https://platform.claude.com/docs/en/agents-and-tools/tool-use/overview}",
    year         = "2026",
    note         = "Accessed 24 August 2026"
}

@misc{GoogleAntigravity,
    author       = "{Google}",
    title        = "{Google Antigravity: Model Context Protocol (MCP)}",
    howpublished = "\url{https://antigravity.google/docs/mcp/}",
    year         = "2026",
    note         = "Accessed 24 August 2026"
}

@misc{MCPRegistry,
    author       = "{Model Context Protocol Project}",
    title        = "{The MCP Registry}",
    howpublished = "\url{https://modelcontextprotocol.io/registry/about}",
    year         = "2026",
    note         = "Accessed 21 August 2026"
}

@misc{A2AProtocol,
    author       = "{A2A Protocol Project}",
    title        = "{Agent2Agent Protocol}",
    howpublished = "\url{https://a2a-protocol.org/latest/}",
    year         = "2026",
    note         = "Accessed 21 August 2026"
}

@misc{ACP,
    author       = "{Zed Industries}",
    title        = "{Agent Client Protocol}",
    howpublished = "\url{https://agentclientprotocol.com}",
    year         = "2025",
    note         = "Accessed 21 August 2026"
}

@misc{AGUI,
    author       = "{AG-UI Project}",
    title        = "{Agent User Interaction Protocol}",
    howpublished = "\url{https://docs.ag-ui.com/introduction}",
    year         = "2025",
    note         = "Accessed 21 August 2026"
}

@misc{OpenAPISpec,
    author       = "{OpenAPI Initiative}",
    title        = "{OpenAPI Specification}",
    howpublished = "\url{https://spec.openapis.org/oas/latest.html}",
    year         = "2026",
    note         = "Accessed 21 August 2026"
}

@article{CWL,
    author  = {Crusoe, Michael R. and Abeln, Sanne and Iosup, Alexandru and Amstutz, Peter and Chilton, John and Tijani{\'c}, Neboj\v{s}a and M{\'e}nager, Herv{\'e} and Soiland-Reyes, Stian and Gavrilovi{\'c}, Bogdan and Goble, Carole and {The CWL Community}},
    title   = {Methods Included: Standardizing Computational Reuse and Portability with the {Common Workflow Language}},
    journal = {Commun. ACM},
    volume  = {65},
    number  = {6},
    pages   = {54--63},
    year    = {2022},
    doi     = {10.1145/3486897}
}

@article{Belov:2010xm,
    author = "Belov, S. and Dudko, L. and Kekelidze, D. and Sherstnev, A.",
    title = "{HepML, an XML-based format for describing simulated data in high energy physics}",
    eprint = "1001.2576",
    archivePrefix = "arXiv",
    primaryClass = "hep-ph",
    doi = "10.1016/j.cpc.2010.06.026",
    journal = "Comput. Phys. Commun.",
    volume = "181",
    pages = "1758--1768",
    year = "2010"
}

@article{Khoo:2022pja,
    author = "Khoo, T. J. and others",
    title = "{Constraints on Future Analysis Metadata Systems in High Energy Physics}",
    eprint = "2203.00463",
    archivePrefix = "arXiv",
    primaryClass = "hep-ex",
    reportNumber = "FERMILAB-PUB-22-345-OCIO-SCD",
    doi = "10.1007/s41781-022-00086-2",
    journal = "Comput. Softw. Big Sci.",
    volume = "6",
    number = "1",
    pages = "13",
    year = "2022"
}

@article{Kasalica:2021ape,
    author  = {Kasalica, Vedran and Schw{\"a}mmle, Veit and Palmblad, Magnus and Ison, Jon and Lamprecht, Anna-Lena},
    title   = {{APE} in the Wild: Automated Exploration of Proteomics Workflows in the bio.tools Registry},
    journal = {J. Proteome Res.},
    volume  = {20},
    number  = {4},
    pages   = {2157--2165},
    year    = {2021},
    doi     = {10.1021/acs.jproteome.0c00983}
}

@article{REANA,
    author = "{\v{S}}imko, Tibor and Heinrich, Lukas and Hirvonsalo, Harri and Kousidis, Dinos and Rodr{\'\i}guez, Diego",
    editor = "Forti, A. and Betev, L. and Litmaath, M. and Smirnova, O. and Hristov, P.",
    title = "{REANA: A System for Reusable Research Data Analyses}",
    reportNumber = "CERN-IT-2018-003",
    doi = "10.1051/epjconf/201921406034",
    journal = "EPJ Web Conf.",
    volume = "214",
    pages = "06034",
    year = "2019"
}

@article{ROCrate,
    author  = {Soiland-Reyes, Stian and Sefton, Peter and Crosas, Merc{\`e} and Castro, Leyla Jael and Coppens, Frederik and Fern{\'a}ndez, Jos{\'e} M. and Garijo, Daniel and Gr{\"u}ning, Bj{\"o}rn and La Rosa, Marco and Leo, Simone and {\'O} Carrag{\'a}in, Eoghan and Portier, Marc and Trisovic, Ana and {RO-Crate Community} and Groth, Paul and Goble, Carole},
    title   = {Packaging Research Artefacts with {RO-Crate}},
    journal = {Data Sci.},
    volume  = {5},
    number  = {2},
    pages   = {97--138},
    year    = {2022},
    doi     = {10.3233/DS-210053}
}

@techreport{W3CPROV,
    author      = {Lebo, Timothy and Sahoo, Satya and McGuinness, Deborah and others},
    title       = {{PROV-O}: The {PROV} Ontology},
    institution = {World Wide Web Consortium},
    type        = {{W3C} Recommendation},
    year        = {2013},
    url         = {https://www.w3.org/TR/prov-o/},
    note        = {Accessed 24 August 2026}
}

@article{CERNAnalysisPreservation,
    author = "Fokianos, Pamfilos and others",
    editor = "Doglioni, C. and Kim, D. and Stewart, G. A. and Silvestris, L. and Jackson, P. and Kamleh, W.",
    title = "{CERN Analysis Preservation and Reuse Framework: FAIR research data services for LHC experiments}",
    doi = "10.1051/epjconf/202024506011",
    journal = "EPJ Web Conf.",
    volume = "245",
    pages = "06011",
    year = "2020"
}

@misc{ClaudeCode,
    author       = "{Anthropic}",
    title        = "{Claude Code}",
    howpublished = "\url{https://github.com/anthropics/claude-code}",
    year         = "2026",
    note         = "Software, version 2.1.235; MCP integration documented at \url{https://code.claude.com/docs/en/mcp}, accessed 18 August 2026"
}

@misc{OpenAICodex,
    author       = "{OpenAI}",
    title        = "{Codex}",
    howpublished = "\url{https://github.com/openai/codex}",
    year         = "2026",
    note         = "Software, version 0.148.0, accessed 18 August 2026"
}

@misc{OpenCode,
    author       = "{Anomaly}",
    title        = "{OpenCode}",
    howpublished = "\url{https://github.com/anomalyco/opencode}",
    year         = "2026",
    note         = "Software, version 1.18.18; MCP integration documented at \url{https://opencode.ai/docs/mcp-servers/}, accessed 18 August 2026"
}

@misc{willard2023efficientguidedgenerationlarge,
      title={Efficient Guided Generation for Large Language Models}, 
      author={Brandon T. Willard and Rémi Louf},
      year={2023},
      eprint={2307.09702},
      archivePrefix={arXiv},
      primaryClass={cs.CL},
      url={https://arxiv.org/abs/2307.09702}, 
}

@misc{AISuite,
    author       = {Ng, Andrew and contributors},
    title        = {{aisuite}: Simple, Unified Interface to Multiple Generative {AI} Providers},
    howpublished = {\url{https://github.com/andrewyng/aisuite}},
    year         = {2024},
    note         = {Software, version 0.1.14, accessed 23 June 2026}
}

@misc{Roman:2026rcn,
    author = "Roman, Alexander and Roman, Jacob",
    title = "{Orchestral AI: A Framework for Agent Orchestration}",
    eprint = "2601.02577",
    archivePrefix = "arXiv",
    primaryClass = "cs.AI",
    month = "1",
    year = "2026"
}

@article{SoftwareCitationPrinciples,
    author  = {Smith, Arfon M. and Katz, Daniel S. and Niemeyer, Kyle E. and {FORCE11 Software Citation Working Group}},
    title   = {Software Citation Principles},
    journal = {PeerJ Comput. Sci.},
    volume  = {2},
    pages   = {e86},
    year    = {2016},
    doi     = {10.7717/peerj-cs.86}
}

@misc{toolbase,
    author       = {Menzo, Tony and Roman, Alexander},
    title        = {{\textsc{ToolBase}}},
    howpublished = {\url{https://github.com/alexr314/toolbase}},
    year         = {2026},
    note         = {Accessed August 2026}
}

@misc{Menzo:inprep,
    author = {Menzo, Tony and Fleming, George T. and Matchev, Konstantin T. and Mrenna, Stephen and Roman, Alexander},
    note   = {In preparation}
}

@misc{wang2024toolsanywaysurveylanguage,
      title={What Are Tools Anyway? A Survey from the Language Model Perspective}, 
      author={Zhiruo Wang and Zhoujun Cheng and Hao Zhu and Daniel Fried and Graham Neubig},
      year={2024},
      eprint={2403.15452},
      archivePrefix={arXiv},
      primaryClass={cs.CL},
      url={https://arxiv.org/abs/2403.15452}, 
}

@misc{schick2023toolformerlanguagemodelsteach,
      title={Toolformer: Language Models Can Teach Themselves to Use Tools}, 
      author={Timo Schick and Jane Dwivedi-Yu and Roberto Dessì and Roberta Raileanu and Maria Lomeli and Luke Zettlemoyer and Nicola Cancedda and Thomas Scialom},
      year={2023},
      eprint={2302.04761},
      archivePrefix={arXiv},
      primaryClass={cs.CL},
      url={https://arxiv.org/abs/2302.04761}, 
}

@misc{qin2023toolllmfacilitatinglargelanguage,
      title={ToolLLM: Facilitating Large Language Models to Master 16000+ Real-world APIs}, 
      author={Yujia Qin and Shihao Liang and Yining Ye and Kunlun Zhu and Lan Yan and Yaxi Lu and Yankai Lin and Xin Cong and Xiangru Tang and Bill Qian and Sihan Zhao and Lauren Hong and Runchu Tian and Ruobing Xie and Jie Zhou and Mark Gerstein and Dahai Li and Zhiyuan Liu and Maosong Sun},
      year={2023},
      eprint={2307.16789},
      archivePrefix={arXiv},
      primaryClass={cs.AI},
      url={https://arxiv.org/abs/2307.16789}, 
}

@article{Allanach:2008qq,
    author = "Allanach, B. C. and others",
    title = "{SUSY Les Houches Accord 2}",
    eprint = "0801.0045",
    archivePrefix = "arXiv",
    primaryClass = "hep-ph",
    reportNumber = "FERMILAB-PUB-07-036-T, SLAC-PUB-12765, CERN-PH-TH-2007-148, DAMTP-2007-76, EDINBURGH-2007-31, KEK-TH-1170, LAPTH-1204-07, LPT-ORSAY-07-81, SHEP-07-13",
    doi = "10.1016/j.cpc.2008.08.004",
    journal = "Comput. Phys. Commun.",
    volume = "180",
    pages = "8--25",
    year = "2009"
}

@article{Alwall:2006yp,
    author = "Alwall, J. and others",
    title = "{A Standard format for Les Houches event files}",
    eprint = "hep-ph/0609017",
    archivePrefix = "arXiv",
    reportNumber = "FERMILAB-PUB-06-337-T, CERN-LCGAPP-2006-03",
    doi = "10.1016/j.cpc.2006.11.010",
    journal = "Comput. Phys. Commun.",
    volume = "176",
    pages = "300--304",
    year = "2007"
}

@article{Buckley:2019xhk,
    author = {Buckley, Andy and Ilten, Philip and Konstantinov, Dmitri and L{\"o}nnblad, Leif and Monk, James and Pokorski, Witold and Przedzinski, Tomasz and Verbytskyi, Andrii},
    title = "{The HepMC3 event record library for Monte Carlo event generators}",
    eprint = "1912.08005",
    archivePrefix = "arXiv",
    primaryClass = "hep-ph",
    reportNumber = "MPP-2019-258, MCNET-19-27, LU-TP 19-58",
    doi = "10.1016/j.cpc.2020.107310",
    journal = "Comput. Phys. Commun.",
    volume = "260",
    pages = "107310",
    year = "2021"
}

\clearpage
\onecolumngrid
\appendix
\section{Glossary of Technical Terms}
\label{app:glossary}

The vocabulary of agentic systems is drawn largely from software engineering and machine learning. This appendix collects the terms used in the main text, defines them as they are used in this Letter, and, where a useful correspondence exists, notes the analogous construct in the computational practice of the field. Terms defined elsewhere in the glossary are set in \emph{italics} at their first appearance within an entry.

\begingroup
\small
\interlinepenalty=10000
\newcommand{\hepglossaryentry}[1]{%
  \item[]\hspace*{-1.5em}\textbf{#1}\par\nopagebreak[4]\smallskip%
}
\begin{list}{}{%
  \setlength{\leftmargin}{1.5em}%
  \setlength{\rightmargin}{0pt}%
  \setlength{\labelwidth}{0pt}%
  \setlength{\labelsep}{0pt}%
  \setlength{\itemindent}{0pt}%
  \setlength{\itemsep}{0.7\baselineskip}%
  \setlength{\parsep}{0pt}%
  \setlength{\topsep}{0.5\baselineskip}%
}

\hepglossaryentry{Adapter (wrapper)}
A simple bridge 
between an existing interface and a new
one connected to the model that
does not require a reimplementation of the underlying computation. If the native software already validates inputs, returns structured output, and reports actionable errors, a \emph{tool} can be an adapter, and the scientific implementation and its validation stay in the underlying code.

\hepglossaryentry{Agent}
An LLM placed in an executable environment such that its generated output can trigger actions---running code, invoking external programs, reading files---whose results re-enter its \emph{context} on the next \emph{turn}. The distinguishing feature relative to a chat interface is the closed loop between generation and execution.

\hepglossaryentry{Application programming interface (API)}
The set of functions, classes, and data structures a software library exposes to calling code. When an \emph{agent} uses a library through its native API, the model must generate syntactically valid code against it, which requires either inspecting the codebase or reconstructing the interface from memory.

\hepglossaryentry{Artifact}
Any durable object produced during a run---an event file, a plot, a fitted result, a generated script, a log. Artifacts are the objects with \emph{provenance} that post-hoc grading inspects.

\hepglossaryentry{Benchmark}
A shared set of tasks with a defined grading procedure, used to compare systems. The limitation identified in the main text is not task realism but under-specification: reported configurations do not resolve every component needed to vary models, providers, \emph{harnesses}, and \emph{loadouts} independently.

\hepglossaryentry{Bundle}
A named group of tools within a \emph{toolkit} that together provide a coherent capability. A bundle allows a related subset of tools to be selected into a \emph{loadout} without selecting the entire toolkit or enumerating each tool individually.

\hepglossaryentry{Benchmark condition}
The full set of separately specified and resolved factors defining one measurement: the scientific task, \emph{task scaffolding}, model and inference controls, inference provider, \emph{harness}, and \emph{capability loadout}. Resolving them separately allows a validated operation to be held fixed while other components are varied---the agentic counterpart of varying one generator setting at a time.

\hepglossaryentry{Command-line interface (CLI)}
The \emph{invocation surface} of a program as run from a shell: its executable name, subcommands, flags, and arguments. A common target for \emph{fixed backend} systems, since a generated command is short and directly loggable.

\hepglossaryentry{Compatibility resolver}
The proposed checker that consumes \emph{contracts}: it verifies global constraints when a \emph{loadout} is assembled and compares required against provided properties at each handoff between operations, classifying the composition as compatible, convertible through a registered \emph{adapter}, incompatible, or unresolved. Properties fixed by invocation arguments attach to the resulting \emph{artifact} and are re-checked downstream. The full contracts can stay outside the model's \emph{context}; only mismatches, available conversions, and unresolved assumptions need be surfaced to the agent and the researcher.

\hepglossaryentry{Context (context window)}
The bounded sequence of tokens visible to the model on a given forward pass: system instructions, conversation history, retrieved documents, \emph{tool} descriptions, and tool results. It is a finite resource. The observation that ``descriptions consume context'' is the operational reason that exposing every available capability to every agent does not scale, and hence the motivation for \emph{loadouts}.

\hepglossaryentry{Decoding / sampling policy}
The rule by which tokens are drawn from the model's predicted distribution: greedy (argmax), or stochastic with temperature and nucleus (top-$p$) truncation. Because decoding is generally stochastic, a fixed prompt and a fixed model define a distribution over outcomes rather than a single outcome, which is why repeated trials and a reported sampling policy are required for any measurement (Sec.~II of the main text).

\hepglossaryentry{Dependency / environment}
The full set of software an operation requires in order to run, together with the mechanism that reproduces it (lockfile, container image, environment specification). Physics results generally depend on the whole environment, not on the top-level package alone.

\hepglossaryentry{Directed acyclic graph (DAG)}
A workflow representation in which nodes are operations and edges are dependencies, with no cycles. Used both to fix the order of stages in a workflow-constrained system and to record the dependency structure of a completed run.

\hepglossaryentry{Distribution / package}
An installable unit bundling an implementation with its declared dependencies, so that a capability can be obtained and run by someone other than its author without manual reassembly.

\hepglossaryentry{Domain-specific language (DSL)}
A small purpose-built language whose grammar admits only meaningful constructions in the target domain, in contrast to a general-purpose language. Restricting generation to a DSL narrows the space of expressible programs, and hence of possible errors.

\hepglossaryentry{Fixed backend}
An architecture in which the \emph{agent} does not implement the physics at all but generates the input consumed by established software: a command-line invocation, a run card or configuration file, a template, or a \emph{DSL} program. Validation then rests on the underlying program.

\hepglossaryentry{Free-form (improvisational, token-by-token) generation}
An \emph{interaction} surface in which the \emph{agent} writes arbitrary code and shell commands to accomplish the task, with domain knowledge supplied only as \emph{context}. Maximally flexible, and the appropriate mode while a method is still evolving; every generated implementation must, however, be independently verified.

\hepglossaryentry{Guardrail}
Any programmatic constraint restricting what an \emph{agent} may do, i.e.~filesystem or network restrictions, forbidden commands, required approval before an irreversible action. Distinct from an \emph{oversight policy}, which specifies where a human must intervene.

\hepglossaryentry{Harness}
The software surrounding the model that turns a general-purpose LLM into a domain \emph{agent}: the policies for context management, persistent state, execution, retry, stopping, error recovery, and logging, together with the interfaces through which the model acts. 
The \emph{harness} is a separately identifiable component from the model, and a measured change in performance cannot be attributed without holding it fixed.

\hepglossaryentry{Harness engineering}
The design discipline of deciding what the \emph{harness} fixes and what it leaves to the model: which operations are promoted to validated interfaces, how they are presented, how failures are surfaced and recovered, and where human verification remains mandatory.

\hepglossaryentry{Interface conformance vs. scientific validity}
The central distinction of Sec.~I: schema mechanisms of either kind establish only that a call is well-formed. Whether the answer is correct depends on the implementation, its validation record, its version, its declared conventions, and whether it was composed correctly with the operations around it.

\hepglossaryentry{Invocation surface}

The way a program is invoked from a shell -- the executable name, subcommands, flags, and arguments.

\hepglossaryentry{Leakage / contamination}
Availability of the answer to the system by an unintended route---presence in the model's training data, reachability from the sandbox, or retrieval of the source publication. Public/private task splits and unpublished problems are the standard mitigations.

\hepglossaryentry{Loadout}
A version-resolved selection of toolkits, bundles, individual \emph{tools}, and associated \emph{skills} exposed to an agent for a particular task. Loadouts exist because presenting every available capability at once is impractical: descriptions consume \emph{context}, overlapping operations complicate selection, and operations with incompatible conventions may not compose safely. A registered loadout also serves as executable documentation of a published workflow.

\hepglossaryentry{Long-horizon task}
A task requiring many dependent \emph{turns}, where errors compound and intermediate results must be carried forward. The characteristic failure mode is not a single wrong answer but a silent early mistake propagated through subsequent stages, which is what motivates stagewise checks.

\hepglossaryentry{Machine-readable scientific contract}
Proposed in Sec.~II as the layer that interface schemas do not provide: a declaration, in namespaced and versioned vocabulary, of the scientific properties an operation \emph{requires} of its inputs and \emph{guarantees} for its outputs---units, reference frame, event-record and status conventions, weight normalization, perturbative order, PDF and tune choices, object definitions, statistical conventions, domain of validity. A schema says an argument is a float; a contract says it is a mass in GeV in the laboratory frame at a stated order.

\hepglossaryentry{Matched / factorial comparison}
An experimental design varying one factor of the \emph{benchmark condition} at a time (matched), or crossing several systematically (factorial), so that an observed change can be attributed to a component. 
This is the reason for the separate resolution of condition factors.

\hepglossaryentry{MCP server / client / host}
The server exposes operations and executes them, dispatching each either directly or onto an existing library API, CLI program, or DSL; the client, inside the \emph{harness}, discovers and calls them; the host is the application the user interacts with. Related protocols address communication between independently deployed \emph{agents}, and between agents and user interfaces.

\hepglossaryentry{Metadata}
Structured descriptive information carried alongside data or code, such as units, conventions, generator settings, selection definitions. Existing HEP metadata efforts demonstrate both the value of recording such information and the difficulty of maintaining it across heterogeneous analysis systems.

\hepglossaryentry{Model Context Protocol (MCP)}
An open protocol standardizing how capabilities are discovered, described, invoked, and returned between a \emph{harness} (the client) and a capability provider (the server). Its role here is portability: a developer exposes a scientific operation once as an MCP server, and any MCP-capable harness translates its description and calls into whatever format the active inference provider expects. The role played is analogous to that of an interchange format such as the Les Houches Event file;
it is not a computation, but an agreement about how independently developed codes exchange it.

\hepglossaryentry{Namespaced, versioned declaration}
A property name qualified by the vocabulary that defines it and the version of that vocabulary, so that independently developed operations do not collide on a shared word or silently drift in what it means.

\hepglossaryentry{Oversight policy}
The specification of where human review is required, what the reviewer may see, and what interventions are permitted. Because interventions change the \emph{trajectory}, they must be declared as part of the condition and recorded in the \emph{trial record} rather than left implicit.

\hepglossaryentry{Pinning / resolution}
Pinning declares an exact version rather than a range; resolution is the process of turning a set of declared requirements into a concrete set of installed versions. The distinction between the \emph{requested} and \emph{resolved} configuration is why both must appear in a \emph{trial record}: the former is the intent, the latter is what actually ran.

\hepglossaryentry{Post-hoc grading}
Evaluating or re-evaluating a retained \emph{trajectory} and its \emph{artifacts} after the fact, for example under a revised rubric, without re-running the model. It requires that the trial record be complete enough to stand on its own, and it makes each reported score dependent on a versioned rubric that should be identified alongside it.

\hepglossaryentry{Provenance}
The recorded lineage of a result: which operations produced it, with which arguments, at which versions, in which environment. An agentic workflow without provenance can produce a number that cannot be traced to the computation that generated it.

\hepglossaryentry{Registry}
An index of available capabilities recording, for each, the \emph{metadata} needed to resolve it to a fixed implementation, agent-facing interface, dependency set, and citable version. It provides the discovery layer: the mechanism that enables a researcher, or an agent, to find a preexistent operation and not need to reimplement it.

\hepglossaryentry{Reproducibility (as used here)}
Obtaining the same result from the same declared condition, realized environment, and retained evidence. It is a property of what was preserved, not only of what was reported: a score without its resolved configuration and trial record cannot be reproduced or reinterpreted, only quoted.

\hepglossaryentry{Rubric}
The explicit criteria for scoring an outcome. 
A physics-aware rubric grades scientific content, such
as the derivation, the distribution, or the observable,
that is independent of which toolkit or source structure produced it, so that results from different implementations can be compared fairly.

\hepglossaryentry{Sampling policy}
The decoding parameters and the number of repeated trials per condition. Since a condition induces a distribution rather than a point, both must be fixed and reported.

\hepglossaryentry{Sandbox / workspace}
The isolated execution environment in which the agent's commands actually run.  The sandbox is typically a container or virtual machine with a fixed filesystem, installed software, and restricted network access.
Its contents matter for evaluation integrity: ground truth must be kept outside it.

\hepglossaryentry{Scaffolding}
The initialized state supplied to an agent before the task begins: the prepared \emph{sandbox} and workspace, system and task prompts, supplied reference information, and procedural recipes. In the benchmarking framework of the main text, scaffolding is specified and resolved separately from the model, the \emph{harness}, and the \emph{loadout}, so that it can be held fixed while the others are varied.

\hepglossaryentry{Schema}
A machine-readable declaration of the structure a data object must have: field names, types, permitted values, required fields. JSON Schema is the format in which tool argument structures are conventionally declared. A schema constrains \emph{form}, and nothing more.

\hepglossaryentry{Semantic versioning}
The convention of numbering releases \texttt{MAJOR.MINOR.PATCH}, where a MAJOR increment signals a backwards-incompatible interface change. It provides the mechanical basis for a \emph{compatability resolver} to decide whether two declared requirements can be satisfied simultaneously.

\hepglossaryentry{Skill}
A packaged unit of procedural knowledge
loaded into the agent's \emph{context} rather than executed.
Examples include when to use which operations, in what order, and with what checks and conventions. 
While a \emph{tool} queries what can be called, a skill answers how the calls should be sequenced and interpreted. The nearest familiar analogue is a well-written recipe or analysis note that accompanies a code release.

\hepglossaryentry{Stagewise check}
Verification applied at intermediate points of a \emph{long-horizon} workflow, such as assumptions, intermediate derivations, simulation and data-processing stages, and final observables, so that a failure is localized rather than observed only as a wrong final answer.

\hepglossaryentry{Structured handoff}
An architecture that leaves scientific operations to generated code but constrains the \emph{intermediate artifacts} passed between \emph{agents} or workflow 
stages.  Examples are fixed JSON schemas, machine-readable project specifications, or a prescribed selection list. It constrains the interfaces between stages rather than the computations within them.

\hepglossaryentry{Tool}
Following the LLM tool-use literature and as used throughout this Letter, a function interface through which the model invokes a program external to itself by generating a call and its arguments. 
A tool definition presented to the model consists of a name, a natural-language description of what the operation does, and a machine-readable \emph{schema} for its arguments. 
Note that ``tool'' is used more loosely elsewhere in the HEP literature, where it may denote a handbook, a command template, a CLI utility, a \emph{DSL}, or a packaged \emph{skill}. Versioning, validation, and schema-constrained generation are additional properties 
of a tool, not part of its definition.

\hepglossaryentry{Tool call}
A single structured invocation emitted by the model: the tool name plus an argument object. The \emph{harness} parses it, dispatches it to the underlying implementation, and returns the result into \emph{context}. Because the call is structured rather than free text, it can be logged, replayed, and resolved to a specific implementation version.

\hepglossaryentry{Toolkit}
The set of callable operations that a given body of software makes available as \emph{tools}, defining the vocabulary the model can select from and compose.

\hepglossaryentry{Trajectory}
The complete ordered record of one agent run: prompts, model outputs, \emph{tool calls} and their arguments, returned results, errors, retries, human interventions, and produced \emph{artifacts}. The trajectory is the observable of an agentic experiment, in the same sense that an event record is the observable of a generator run; grading is a function applied to it.

\hepglossaryentry{Trial record}
The retained evidence from a single run: requested and resolved configuration, prompts, interface descriptions, \emph{trajectory}, researcher interventions, \emph{artifacts}, software versions, grading evidence, termination status, and resource use. Its purpose is to allow a scientific failure to be distinguished from an infrastructure failure, and to allow re-grading without repeating inference.

\hepglossaryentry{Turn}
One cycle of model generation followed by execution of any requested actions and the return of their results into \emph{context}. A multi-turn interaction is the basic unit of agentic behavior.

\hepglossaryentry{Typed schema}
A \emph{schema} with fields that carry declared types (and often units, ranges, or enumerated values) rather than free strings, so that malformed arguments are detectable mechanically.

\hepglossaryentry{Valid-by-construction (constrained decoding)}
Validation \emph{during} generation: at each step the sampler's support is restricted to tokens consistent with the \emph{schema} or grammar, so a malformed call cannot be emitted at all. Requires support from the inference provider. This is the mechanism placing HEPTAPOD in the corresponding row of Table~I.

\hepglossaryentry{Valid-by-rejection}
Validation \emph{after} generation: the model emits a call, the harness checks it against the \emph{schema}, and invalid structure or values are rejected with diagnostics that re-enter \emph{context} so the model can correct itself. Simple to implement and provider-independent, at the cost of wasted turns.

\hepglossaryentry{Value-level check}
A check on the \emph{content} rather than the form of a call.  Examples are checking that a mass is non-negative, that a final state conserves charge, or that a requested order is within the validity of the implementation. Necessary because a call can be perfectly well-formed and physically meaningless.

\hepglossaryentry{Versioned scientific operation}
The central construct of this Letter: a scientific operation for which the implementation, the agent-facing interface, and any external software it invokes all resolve to fixed versions, with explicit dependencies and a citable identifier. It is the agentic analogue of a tagged, archived, citable code release, and it is what makes an operation reusable across harnesses rather than bound to the system that first demonstrated it.

\end{list}
\endgroup

\end{document}